\documentclass[12pt]{elsarticle}
\usepackage{lineno}

\usepackage{lineno,hyperref}
\usepackage{tabularx}

\usepackage{graphicx}
\usepackage{mwe}
\usepackage{caption}
\usepackage{subcaption}

\usepackage[numbers]{natbib}

\usepackage{color,soul}

\usepackage[utf8]{inputenc}
\usepackage{amsmath}

\usepackage{booktabs}
\usepackage{float}
\usepackage{setspace}
\usepackage{xcolor}
\usepackage{adjustbox}

\usepackage{xcolor,mdframed}
\usepackage{graphicx}
\usepackage[normalem]{ulem}

\usepackage{hyperref}

\makeatletter
\def\@author#1{\g@addto@macro\elsauthors{\normalsize%
    \def\baselinestretch{1}%
    \upshape\authorsep#1\unskip\textsuperscript{%
      \ifx\@fnmark\@empty\else\unskip\sep\@fnmark\let\sep=,\fi
      \ifx\@corref\@empty\else\unskip\sep\@corref\let\sep=,\fi
      }%
    \def\authorsep{\space and\space}%
    \global\let\@fnmark\@empty
    \global\let\@corref\@empty
    \global\let\sep\@empty}%
    \@eadauthor={#1}
}

\date{}

\def\ps@pprintTitle{%
 \let\@oddhead\@empty
 \let\@evenhead\@empty
 \def\@oddfoot{\centerline{\thepage}}%
 \let\@evenfoot\@oddfoot}

\makeatother

\graphicspath{ {images/} }

\usepackage{comment}

\usepackage[a4paper, total={6in, 8in}]{geometry}

\journal{Journal of Data and Information Science}

\biboptions{authoryear}

\begin{document}


\title{Public Reaction to Scientific Research via Twitter Sentiment Prediction}

\author{Murtuza Shahzad}
\ead{msyed1@niu.edu}
\author{Hamed Alhoori}
\ead{alhoori@niu.edu}

\address{Department of Computer Science, Northern Illinois University, DeKalb, IL, USA}

\begin{frontmatter}

\begin{abstract}

    \textbf{Purpose:} Social media users share their ideas, thoughts, and emotions with other users. However, it is not clear how online users would respond to new research outcomes. This study aims to predict the nature of the emotions expressed by Twitter users toward scientific publications. Additionally, we investigate what features of the research articles help in such prediction. Identifying the sentiments of research articles on social media will help scientists gauge a new societal impact of their research articles.

    \textbf{Design/methodology/approach:} Several tools are used for sentiment analysis, so we applied five sentiment analysis tools to check which are suitable for capturing a tweet’s sentiment value and decided to use NLTK VADER and TextBlob. We segregated the sentiment value into negative, positive, and neutral. We measured the mean and median of tweets’ sentiment value for research articles with more than one tweet. We next built machine learning models to predict the sentiments of tweets related to scientific publications and investigated the essential features that controlled the prediction models. 
    
    \textbf{Findings:} 
    We found that the most important feature in all the models was the sentiment of the research article title followed by the author count. We observed that the tree-based models performed better than other classification models, with Random Forest achieving 89\% accuracy for binary classification and 73\% accuracy for three-label classification.

    \textbf{Research Limitations:} In this research, we used state-of-the-art sentiment analysis libraries. However, these libraries might vary at times in their sentiment prediction behavior. Tweet sentiment may be influenced by a multitude of circumstances and is not always immediately tied to the paper's details. In the future, we intend to broaden the scope of our research by employing word2vec models. 

    \textbf{Practical Implications:} Many studies have focused on understanding the impact of science on scientists or how science communicators can improve their outcomes. Research in this area has relied on fewer and more limited measures, such as citations and user studies with small datasets. There is currently a critical need to find novel methods to quantify and evaluate the broader impact of research. This study will help scientists better comprehend the emotional impact of their work. Additionally, the value of understanding the public’s interest and reactions helps science communicators identify effective ways to engage with the public and build positive connections between scientific communities and the public.

    \textbf{Originality/value:} This study will extend work on public engagement with science, sociology of science, and computational social science. It will enable researchers to identify areas in which there is a gap between public and expert understanding and provide strategies by which this gap can be bridged.  
\end{abstract}


\end{frontmatter}

\section{Introduction}
  	  Microblogging has become ubiquitous, and the scope of the texts posted on microblogging platforms and their efficacy as a means of communication has far exceeded expectations. Social media platforms have become a place where users collaborate, share their ideas and also have conflicts \citep{hansson2019capitalizing,hansson2019crowd}. With 126 million active daily users \citep{Shaban2019-ih}, Twitter is the dominant microblogging platform on which users discuss a breadth of subjects and even play a role in influencing current trends. Users on Twitter post short and often informal messages (tweets) in which they share information and project opinions and sentiments about what is going on in the world. Twitter has been a major platform for sharing scholarly articles, and many researchers have used it to develop various metrics for scholarly articles \citep{haustein2019scholarly}. Other social media platforms like Facebook and Weibo have also been sources to study online users' responses \citep{kou2017one}. Social media platforms have become a hub where users express their opinions and emotions related to multiple fields of interest \citep{chatterjee2019understanding}. Researchers have studied the sentiments and emotions associated with research articles on these platforms \citep{freeman2019shared,freeman2020measuring}.

There is a need to understand the impact of research beyond the traditional scholarly impact (i.e., citations)\citep{le2019citeopinion,KoushaThelwall,noyons2019measuring} such as the impact on economics \citep{Abdul2019}, public policy \citep{kale2017}, social media \citep{alhoori2019}, news outlets \citep{harish2017}, and public understanding of science \citep{harish2018}. In the present study, we subjected a collection of tweets to the process of sentiment analysis, which refers to the contextual mining of texts through which subjective information is identified and extracted \citep{liu2012sentiment}. Such subjective information is essential in many business-related opinion-mining contexts. For scholarly literature, tweets can be analyzed to determine the popularity of a research article, whether it is liked, or even whether anyone has shared and discussed the article online. We analyzed a collection of tweets to identify whether tweets about a given research article were predominantly positive, neutral, or negative. We built machine learning models to predict the nature of the sentiments expressed in tweets about a given research article. We considered several distinct evaluative measures for the machine learning models and found that the tree-based models performed better than the other models in predicting tweet sentiments.

Knowing how social media data can be utilized to learn about a research article’s emotional impact is an interesting quest. Such a study can pave the way for scientists to understand the impact their work could have based on the reactions to previous studies. In addition, they would know about the specific features in their paper that could be modified to avoid a negative reaction. To understand and analyze this impact, we consider the following research questions:

\begin{enumerate}

\item 
 
\textbf{RQ1.} Can we build machine learning models to predict the tweet sentiment for research articles?
\item \textbf{RQ2.} What are the crucial features that help in accurately predicting the sentiment of the tweet?

\end{enumerate}

In summary, our contributions include
\begin{enumerate}
\item 
Analyzing tweets related to research articles using various social media features and research domains.
\item Understanding the sentiment of tweets for research articles using various state-of-the-art sentiment analysis libraries.
\item Building machine learning models to predict sentiments of tweets related to scientific publications. 
\end{enumerate}

\section{Related Work}

Twitter is one of the social media platforms on which many sentiment analysis and predictive models are built \citep{jaidka2021rural,haunschild2020library, ibrahim2019decoding, Fereshteh2018}. In numerous studies, researchers have endeavored to predict the sentiment of general tweets and tweets related to specific areas or events. Some researchers have analyzed the sentiments of tweets related to scholarly articles and have built models to predict sentiments of tweets for scholarly articles for specific research domains \citep{hassan2020sentiment}. \cite{bharathwaj2019sentiment} predict the positive, negative, and neutral sentiment of tweets used for research articles in Medicine and Psychiatry Disciplines. They manually labeled 1,099 negatives, 2,000 positives, and 8,000 neutral tweets, built several machine learning models, and identified the best model with the Support Vector Machines (SVM) having 91.6\% accuracy.

To estimate the polarity of tweets, \cite{narr2012language} subjected Twitter data to a language-independent sentiment analysis. They collected tweets in several languages and used the Naive Bayes classifier on the n-gram features to classify the sentiments. The mixed four-language unigram had an accuracy of 71.5\%. Similarly, \cite{bae2012sentiment} studied the polarity and sentiments of tweets posted by celebrities with more than a million followers. The researchers performed a lexical sentiment analysis based on which the sentiment score was calculated. According to the different types of correlational analysis, when a celebrity posted a positive tweet, his/her followers did likewise. Hence, based on tweet similarity, it appears that the celebrities influenced their respective followers. \cite{kharde2016sentiment} tested various classification methods such as SVM, Naive Bayes, and Maximum Entropy using a Twitter dataset. The baseline model was the least accurate of all the classification models. Compared with that model, Maximum Entropy and Naive Bayes each returned a higher accuracy. However, the highest accuracy was achieved using SVM with Unigram and Bigram with stop word removal. In addition to these classification methods, the classification algorithm Naive Bayes was run separately as a Unigram Multinomial Bayes and a Multigram Multinomial Bayes \citep{parikh2009sentiment}. The results showed that between Maximum Entropy and Naive Bayes classification models, based on the unique nature of the tweets, Maximum Entropy could not take advantage of the sequenceable features. Hence each of the Naive Bayes classifiers performed overwhelmingly better than the other classification models did.

There were different approaches and methods applied in various other studies on Twitter. \cite{zaman2010predicting} suggested a probabilistic collaborative filter model that predicts future retweets, thereby showing the spread of information. \cite{hao2011visual} also used a visual-based approach compared to the previous text-based approaches. The approaches discussed thus far rely on traditional methods to pre-process data. However, \cite{da2014tweet} suggested that data can be preprocessed instead, using feature hashing in relation to a bag of words and lexicons, which are then fed into the classification algorithms. Comparing this method with the baseline method of bootstrapping an ensemble framework as used by \cite{hassan2013twitter}, the researchers obtained similar accuracy, suggesting that classifier ensembles can be useful in tweet sentiment analysis. On a similar note, \cite{saif2014stopwords} used several methods—Zipf’s law, term-based random sampling, mutual information, and the classic (pre-compiled) approaches—in conjunction with each other to remove the stop-words during preprocessing. In comparison with the baseline model results in which the stop words were not removed, their method reduced the feature space by nearly 65\%, thereby maintaining high performance on classification by decreasing data sparsity up to 0.37\%.

Likewise, \cite{Pak2010-eg} performed a linguistic analysis of tweets collected for their research. After preprocessing the tweets, the researchers extracted features that were then used in the Multinomial Naive Bayes classifier algorithm to predict positive, neutral, and negative sentiments. According to their results, bigrams are better than both unigrams and trigrams in capturing sentiment expression. Instead of considering all the text comprising a tweet, \cite{kouloumpis2011twitter} suggested analyzing only the hashtags present in the tweets to create what they referred to as a hashtagged dataset. On comparing the average accuracy of various features in predicting tweet sentiment, they found hashtags and emoticons to be more useful than part-of-speech features. \cite{wang2011topic} developed that research direction further by analyzing only the hashtags of tweets and also boosting the graph-based classification algorithms, Loopy Belief Propagation (LBP), Relaxation Labeling (RL), and the Iterative Classification Algorithm (ICA). The researchers compared their model with the baseline model, which relied on traditional classification algorithms. Their results show that boosting achieved better results for predicting positive and negative tweets based on hashtags in terms of accuracy, precision, and recall compared with plain classification algorithms.

\cite{wang2012automatic} used Semantic Role Labeling (SRL) to extract event-based tweets. The researchers used Latent Dirichlet Allocation (LDA) to identify the salient topics in the events. Using these topics, they built a model to predict future criminal events. In a similar attempt to predict crimes, \cite{chen2015crime} used weather as a feature in addition to Twitter sentiment to predict the times and locations of crimes. Based on Kernel Density Estimation (KDE) used in conjunction with lexicon-based methods, they observed that temperature, aggression, and crime rate are related: high temperatures were associated with a high level of aggression, which ultimately led to a higher crime rate than when temperatures were low. Furthermore, in the context of weather, \cite{mandel2012demographic} considered tweets selected based on the level of concern expressed and demographic information related to Hurricane Irene. The researchers found that the number of tweets related to the hurricane directly affected the region peaks at the time of the hurricane and that the level of concern expressed in the tweets depended on the particular region.

In addition to crime and weather, the literature includes analyses of many other topics ranging from the incidence of disease to the stock market. For example, \cite{achrekar2011predicting} analyzed the content of tweets in an endeavor to predict flu trends. In relation to the stock market, \cite{mittal2012stock} drew on Twitter sentiments to analyze stock market trends. The researchers subjected the tweets to natural language processing and then to several classification models. The researchers found that Self Organizing Fuzzy Neural Networks returned the most accurate results.

The literature includes several studies indicating that tweets can influence election results. \cite{pal2018speaking} show how politicians may benefit from antagonistic messaging. \cite{wang2012system} used a real-time data-processing infrastructure on IBM’s InfoSphere stream platform to write visualization modules and perform an analysis of the tweets. In a special case study, \cite{bermingham2011using} analyzed the relationship between emotions expressed in tweets and election results for the Irish general election of 2011. Of the studies published to date, this research is unique in that it differentiates between the emotions expressed in tweets pertaining to the Inter-Party polls and the emotions expressed in tweets pertaining to the Intra-Party polls by implementing a separate measure for each. According to the results, during the weeks before the election, the sentiments were close to each other, but the sentiments were polarized on the day before the election. A model developed by \cite{gayo2012no} predicted that not only would Barack Obama win the US Presidential Election of 2008 but that he would also win all the states. Overall, given that Obama both won the election and all the states with the exception of Texas, the results of the model were very accurate. However, the Texas election result also indicated that such models should be used with caution, given that the tweet sentiments cannot entirely predict the outcome of an election.

According to several research studies, when used with sentiment analysis and the addition of natural language processing, machine learning models can generate relatively accurate results. Exploring further, \cite{neethu2013sentiment} used SVM to train the machine to calculate and predict the sentiment scores of an electronic product. Focusing on movie reviews, \cite{Amolik2016-gc} used machine learning and sentiment scores via a unigram approach to preprocessing the data. The researchers also ensured that hashtags were removed and then stored in the feature vectors. They used both SVM and Naive Bayes to predict the sentiments of movie reviews and then compared the scores to the baseline model on the metrics of precision and recall.

In this paper, we built predictive models to determine the sentiments expressed about research articles in tweets. There are a number of differences between our study and previous work. For example, we used a large dataset in our study compared with other studies. We use tweets related to research articles belonging to a broader range of scientific domains to observe what domains are better indicators of sentiment prediction. Additionally, we did not manually label the sentiment of tweets. Instead, we analyzed different sentiment analysis libraries to obtain class labels that were used as dependent variables in our machine learning models. For class imbalance issues among positive, negative, and neutral sentiments, we used state of the art class imbalance technique called Synthetic Minority Oversampling Technique (SMOTE). Further, some studies have either eliminated or included neutral sentiments for predictions. We built models with and without neutral sentiments to observe how the predictions work. 

\section{Data}
We used Altmetrics data released from Altmetric.com in July 2018. Altmetrics \citep{akella2021early,alhoori2014a, alhoori2014b, alhoori2015}, consist of mentions of scholarly articles in online social media, such as Facebook, Twitter, and Wikipedia, and in online reference managers, such as Mendeley. The 2018 release of the Altmetric dataset consists of the details about the online mentions of about 19 million publications. The data comprise details about research articles, including a given article’s title, author(s), and subject(s), as well as tweets about the article and the number of times it has been shared on social media. In this study, we focused on the research articles shared on Twitter to determine the tweets’ sentiments. To meet this goal, we first filtered the entire Altmetric dataset having the details of tweets. With this filtering, we were left with 6,011,003 articles. We took a random sample of 150,000 research articles from the 6 million articles using the Pandas library \citep{mckinney2011pandas}. This random sampling was done by selecting a certain number of articles (150,000) without replacement from the dataset. Finally, we eliminated all records with missing values. The final dataset had altmetrics for 148,712 research articles with a total of 1,941,348 tweets. The entire study was done on these randomly selected articles. To the best of our knowledge, this is one of the largest datasets in such studies. The altmetrics features used in this study for data analysis and for predicting the tweets' sentiments are mentioned in Table \ref{tab:twitter-features}. Some of the features we used, such as the abstract length of the article and the number of followers of the Twitter user, were shown to be important factors in measuring the popularity of articles on Twitter \citep{pandian2019altmetrics}.

\begin{table*}[!htp]
  \caption{Selected features from the Altmetrics dataset.}
  \label{tab:twitter-features}
  \begin{adjustbox}{center}
  \begin{tabular}{ll}
    \toprule
    \textbf{Feature}  & \textbf{Description}\\
    \midrule
    Scopus subject & 
Subject of a research article. \\
 Article title & Title of a research article. \\
Article abstract & Abstract of a research article. \\
Abstract length & Number of words in the abstract of a research paper. \\
Follower count & Number of followers a Twitter user has.  \\
Author count & Number of authors credited on the research article. \\
Tweet & Tweet about a research article. \\

    \bottomrule
  \end{tabular}
  \end{adjustbox}
\end{table*}

\begin{table*}[!ht]
  \caption{Derived features from the dataset.}
  \label{tab:twitter-features-derived}
  \begin{adjustbox}{center}
  \begin{tabular}{lll}
    \toprule
    \textbf{Original feature} & \textbf{Derived feature}  & \textbf{Description}\\
    \midrule
Article title & Title sentiment & Sentiment score of the title of a research article. \\

Article abstract & Abstract sentiment & Sentiment score of a research article abstract. \\

Follower count & Tweet reach & The mean number of followers of each user who  \\
 & & tweeted about the research article (i.e., one article  \\
 & & can be tweeted by many users, who may differ from \\
 & & each other in the number of followers they have). \\

Tweet & Tweet sentiment & Sentiment score of a tweet related to a \\
& & research article. \\

    \bottomrule
  \end{tabular}
  \end{adjustbox}
\end{table*}

From the features listed in Table \ref{tab:twitter-features}, we derived some new features in Table \ref{tab:twitter-features-derived}. The final features used for the machine learning models are title sentiment, abstract sentiment, abstract length, tweet reach, author count, and tweet sentiment. Given that tweet sentiment is the target variable, we developed machine learning models to predict this variable.

\begin{figure}[hbt!]
    \centering
    \includegraphics[width=\textwidth]{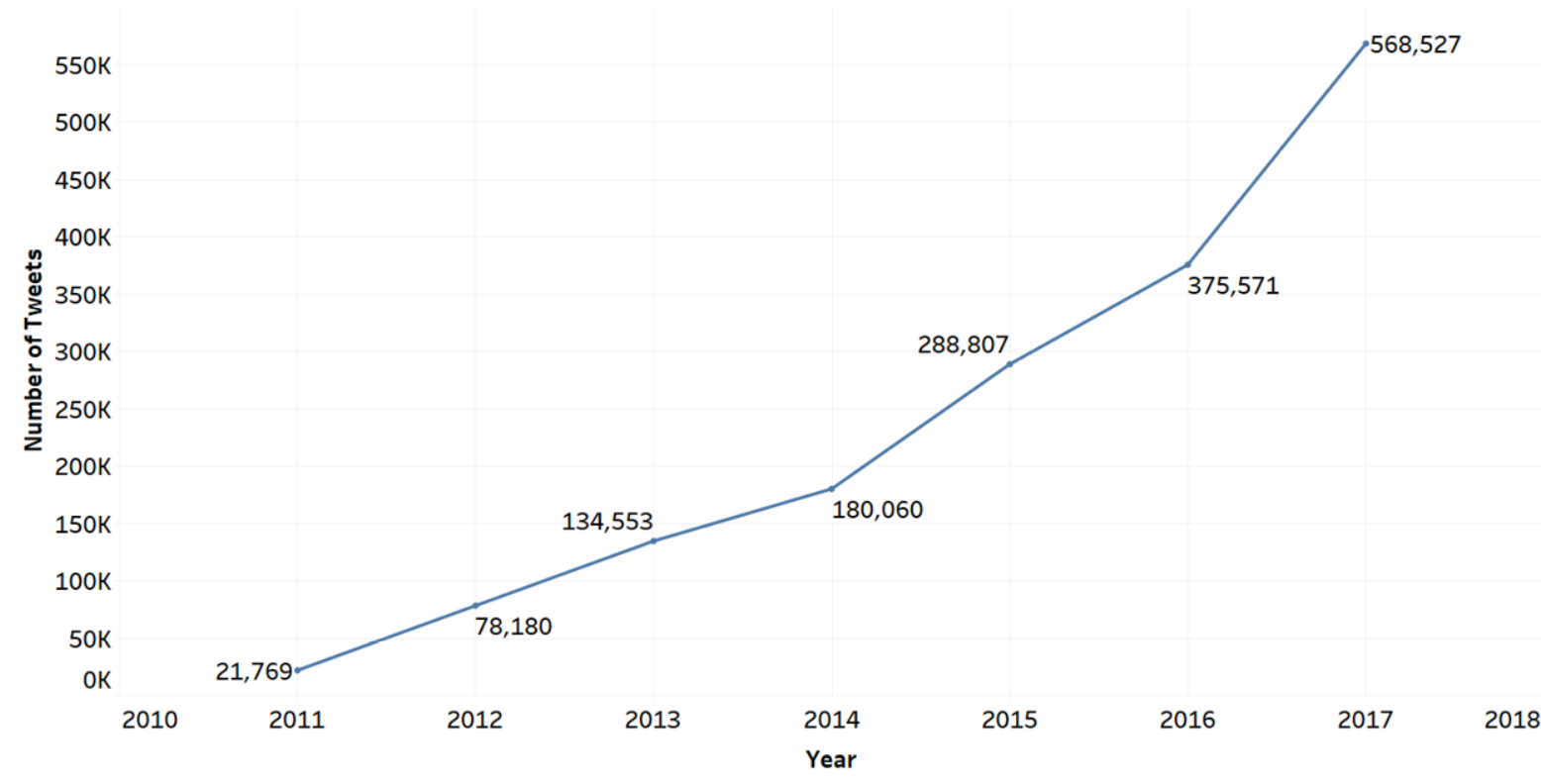}
    \caption{Number of tweets related to research articles for the years 2011-2017.}
    \label{fig:num-of-tweets}
\end{figure}

We used open-source libraries such as Pandas and Matplotlib to load, manipulate, analyze, and visualize the data. We plotted graphs showing various insights. We first observed the number of online mentions of research articles on Twitter from 2011 to 2017. Figure \ref{fig:num-of-tweets} shows an increase in the number of articles shared on Twitter in this timespan. As the dataset did not have the altmetrics for the entire year of 2018, we removed this year from the plot to avoid misinterpretation because of the decrease in tweets for 2018. The number of tweets in the year 2018 (until July) was 293,881.

\begin{figure}[hbt!]
    \centering
    \includegraphics[width=\textwidth]{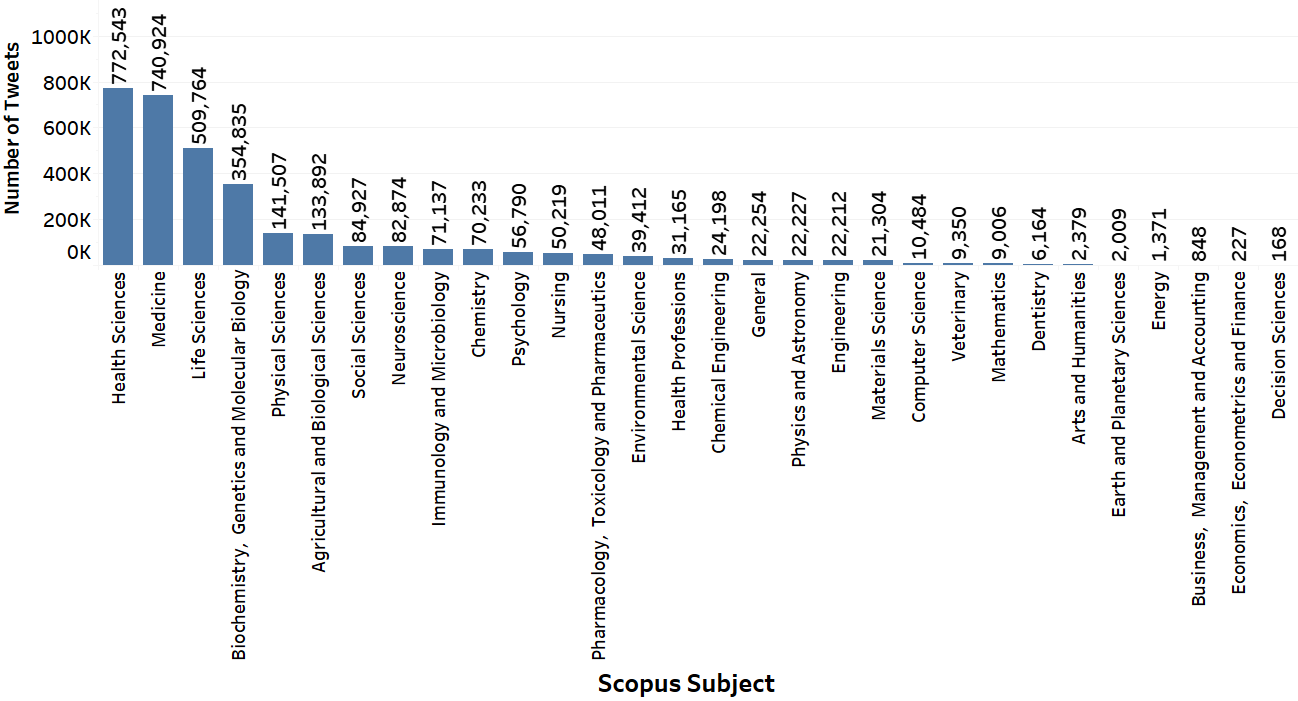}
    \caption{Number of tweets for each Scopus subject.}
    \label{fig:num-tweets-subjects}
\end{figure}

\begin{figure}[hbt!]
    \centering
    \includegraphics[width=\textwidth]{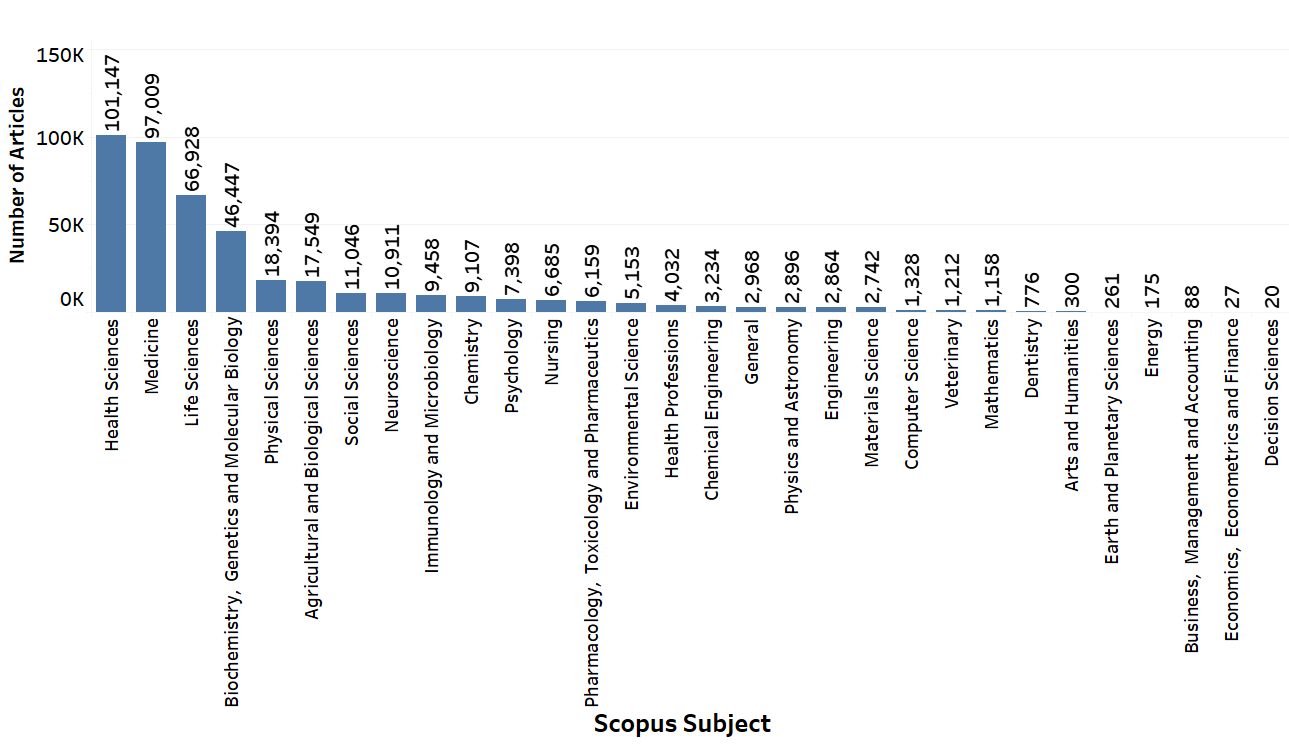}
    \caption{Number of articles for each Scopus subject.}
    \label{fig:num-articles-subjects}
\end{figure}

We then extracted the Scopus subjects from the dataset to determine the extent of their popularity on Twitter. 
Figure \ref{fig:num-tweets-subjects} shows the number of tweets for each subject, and Figure \ref{fig:num-articles-subjects} shows the number of articles for each subject. We observed that Health Sciences had the most articles and most tweets, followed by Medicine.

We obtained 30 Scopus subjects\footnote{ \url{https://service.elsevier.com/app/answers/detail/a\_id/12007/supporthub/scopus/}} from the dataset, which have four higher-level groupings and 27 lower-level groupings of subjects. The four higher-level subjects are Physical Sciences, Health Sciences, Social Sciences, and Life Sciences. It is to be noted that the subject Social Sciences is in the higher and the lower-level groupings.

We noticed that the dataset from Altmetric categorizes Scopus subjects by merging the higher and lower level groupings of Scopus subjects. As an article may belong to multiple Scopus subjects, the Altmetric dataset also categorizes a single article to multiple Scopus subjects. For example, an article in the Altmetric dataset titled ``Respiratory Factors Contributing to Exercise Intolerance in Breast Cancer Survivors: A Case-Control Study'' falls under the category of Medicine, Health Sciences, and Nursing Scopus subjects.

Detecting the sentiment of any given text involves some challenges. \cite{mohammad2017challenges} pointed out that the determination of sentiment could be at different text granularities such as sentiment of a word, a sentence, a paragraph, or an entire document. Mohammad discussed that another challenge is setting a threshold for negative, neutral, and positive sentiments and suggested that some applications may just require the detection of extremely positive and negative instances. \cite{hussein2018survey} found that domain-dependence is another important component in recognizing sentiment. Figures of speech, semantics, explicit and implicit opinions, regular and comparative opinions in tweets are challenges for identifying sentiment analysis of tweets \citep{pozzi2017challenges}. In analyzing the sentiments in academic writing, \cite{vinkers2015use} found that the use of positive words like novel, robust, innovative, and unprecedented in the titles and abstracts of research articles published between 1974 and 2014 grew significantly. Negative words have also increased in frequency, albeit in a smaller but statistically significant way. For the research articles selected in our study, Table \ref{tab:20posnegwords} shows the top 25 positive and negative words in the title, abstract, and tweets related to the articles given by the TextBlob sentiment analysis library.

\begin{table}[hbt!]
\centering
\caption{Top 25 positive and negative words in title, abstract, and tweets of research articles.}
\label{tab:20posnegwords}

\begin{adjustbox}{center}
\begin{tabular}
{|p{2.5cm} |p{2.5cm}| p{2.5cm} |p{2.5cm}| p{2.5cm}| p{2.5cm}| }
\hline
                             \multicolumn{2}{|c|}{\textbf{Title}} & \multicolumn{2}{c|}{\textbf{Abstract}} &
                             \multicolumn{2}{c|}{\textbf{Tweets}} 
                             \\ \hline
                             
                             \textbf{Positive}   &\textbf{Negative}&
                             \textbf{Positive}   &\textbf{Negative}&
                             \textbf{Positive}   &\textbf{Negative}
                             \\ \hline

best &boring        &awesome    &awful        &awesome      &awful      \\ \hline
delicious &devastating        &best    &bleak     &best      &bleak \\ \hline
excellent &disgusting         &delicious    &boring        &breathtaking      &boring      \\ \hline
greatest &evil      &excellent    &cruel        &delicious      &cruel     \\ \hline
perfect &grim         &exquisite   &devastating        &delightful   &devastating   \\ \hline
superb &vicious      &flawless    &disgusted       &excellent      &disgusting      \\ \hline
wonderful &worst        &greatest    &dreadful     &exquisite      &dreadful   \\ \hline
brilliant &fearful         &impressed   &evil        &greatest     &evil      \\ \hline
ideal &repellent       &legendary    &grim        &impressed     &grim      \\ \hline
incredible &retard         &magnificent    &gruesome        &legendary      &gruesome      \\ \hline
beautiful &base         &marvelous    &horrible        &magnificent      &horrible      \\ \hline
splendid &bloody         &masterful    &horrific        &marvelous      &horrific      \\ \hline
attractive &doubtful         &perfect    &hysterical        &masterful     &hysterical   \\ \hline
experienced &filthy         &superb    &insane        &perfect     &insane      \\ \hline
expressive &grief        &wonderful   &insulting       &priceless      &insulting      \\ \hline
favored &hate         &artesian   &menacing        &superb     &miserable      \\ \hline
great &violent          &brilliant    &outrageous        &wonderful      &nasty      \\ \hline
happy &stupid          &ideal   &ruthless        &brilliant      &outrageous      \\ \hline
intelligent &tragic  &incredible    &shocking        &ideal      &pathetic      \\ \hline
joy &sick         &beautiful    &terrible        &incredible      &shocking     \\ \hline
proud &anger           &attractive   &terrifying        &beautiful      &terrible     \\ \hline
uncommon &crude          &brave   &vicious        &splendid      &terrifying      \\ \hline
unforgettable &frustrated           &elect   &worst        &attractive      &vicious      \\ \hline
win &painful           &experienced   &fearful        &brave      &worst      \\ \hline
remarkable &shocked           &expressive   &hated        &elect      &fearful     \\ \hline

\end{tabular}
\end{adjustbox}

\end{table}

\section{Methods}

To obtain sentiment scores for the tweets, the title of the research articles, and the abstracts of the research articles, we used the following Python libraries:

\begin{enumerate}
    \item \textbf{NLTK\footnote{\url{https://www.nltk.org/}} VADER} (Valence Aware Dictionary and sEntiment Reasoner) is a lexical and rule-based sentiment analysis tool in the Natural Language Toolkit library (NLTK). A component of VADER, the sentiment intensity analyzer generates a compound polarity score for text. This score is a continuous value in the range of -1 (negative) to +1 (positive). We used version 3.6.2 of the NLTK library.
    \item \textbf{TextBlob\footnote{\url{https://pypi.org/project/textblob/}}} is a library used for various purposes, such as part-of-speech tagging, noun phrase extraction, and sentiment analysis. This library also generates a sentiment score as a continuous value in the range of -1 (negative) to +1 (positive). We used version 0.15.3 of the TextBlob library.
    \item \textbf{Stanford CoreNLP\footnote{ \url{ https://stanfordnlp.github.io/CoreNLP/}}}  is a library built in Java. For Stanford CoreNLP, the Python packages interact with the library on a server running in the background on the Java platform. This library generates sentiment scores as follows: very negative = 0, negative = 1, neutral = 2, positive = 3, and very positive = 4. We used version 4.2.0 of the Stanford CoreNLP library.
    \item \textbf{SentiStrength\footnote{ \url{ http://sentistrength.wlv.ac.uk/}}}  is an algorithm by \cite{thelwall2010} used to extract sentiments of informal texts. We used a Python wrapper of this algorithm to generate ternary sentiments: negative = -1, neutral = 0, positive = 1. 
    \item \textbf{Sentiment140\footnote{ \url{ http://www.sentiment140.com/}}}  is a sentiment analysis tool specifically designed for Twitter. We used the sentiment140 API to query the tweets’ sentiment and obtain the polarities as negative = 0, neutral = 2, and positive = 4.
\end{enumerate}

Each article might receive multiple tweets (e.g., tweets by different users), and the sentiment of those tweets determine the positive, negative, or neutral impact of the research article. To get an overall sentiment value for a research article, we used the mean and median of multiple tweet sentiments for a particular research article. We looked at the sentiment scores these libraries gave to the tweets. If they gave a high level of neutral sentiments, we did not use those libraries in building machine learning models. The Stanford CoreNLP library generated a score of neutral for most tweets such that we could not use the scores in our study. Table \ref{tab:new-lib-and-sent-twitter} shows that SentiStrength and Sentiment140 libraries resulted in many neutral sentiments. Surprisingly, we observe that the SentiStrength library gave more negative sentiments than positive sentiments. Some studies \citep{friedrich2015tweets,friedrich2015adapting} have shown that applying the SentiStrength library to the tweets of  scientific  articles resulted in a high number of neutral sentiments of the tweets. However, for the purpose of this study, it was appropriate to select those libraries that would generate more non-neutral sentiments. This would help us to build machine learning models that have less class imbalance. Therefore, as the Stanford CoreNLP, SentiStrength, and Sentiment140 libraries generated most of the tweets' sentiment as neutral, we did not use these libraries any further in our study.

\begin{table*}[h!]
  \caption{Sentiment distribution of articles using SentiStrength and Sentiment140 libraries.}
  \label{tab:new-lib-and-sent-twitter}
  \begin{adjustbox}{center}
  \begin{tabular}{lllll}
    \toprule
    
\vtop{
\hbox{\strut Sentiment}
\hbox{\strut library}
} &
\vtop{
\hbox{\strut Metric for}
\hbox{\strut multiple }
\hbox{\strut sentiments}
} & 
\vtop{
\hbox{\strut Number of}
\hbox{\strut positive}
\hbox{\strut sentiments}
} & 
\vtop{
\hbox{\strut Number of}
\hbox{\strut negative}
\hbox{\strut sentiments}
} & 
\vtop{
\hbox{\strut Number of}
\hbox{\strut neutral}
\hbox{\strut sentiments}
}

\\
    \midrule
SentiStrength & mean & 11,443 ($ \approx7.7\%$) & 31,212 ($ \approx21\%$) & 106,057 ($\approx71.3\%$)\\
SentiStrength & median & 14,905 ($ \approx10\%$) & 39,091 ($ \approx26.3\%$) & 94,716 ($ \approx63.7\%$)\\
Sentiment140 & mean & 3,528 ($ \approx2.4\%$) & 6,254 ($ \approx4.2\%$) & 138,930 ($  \approx93.4\%$)\\
Sentiment140 & median & 3,544($ \approx2.4\%$) & 3,168 ($ \approx2.1\%$) & 142,000 ($ \approx95.5\%$)\\
    \bottomrule
  \end{tabular}
  \end{adjustbox}
\end{table*}

All the data were numeric, ranging from zero to tens of thousands. Thus, to convert this sparse data into meaningful machine-interpretable data, we applied a feature-scaling technique. We used a standardization methodology (Z-score normalization) whereby the features were rescaled so they would have the properties of a standard normal distribution. We built three kinds of machine learning models broadly classified as follows:
\begin{enumerate}
\item Classification models to predict the tweet sentiments as binary class labels (positive and negative)
\item  Classification models to predict the tweet sentiments as one of three class labels (positive, neutral, and negative)
\item Regression models to predict the exact tweet sentiment scores.
\end{enumerate}

As we built the machine learning models for a target variable that was purely based on the sentiment analysis libraries, it was important to verify that the sentiment score generated using these libraries were reliable. For this purpose, two individuals manually labeled a random set of 200 tweets for positive and negative sentiment. For the manually labeled sentiments, we evaluated the Cohen's kappa coefficient \citep{cohen1960coefficient}, which is a statistic for determining inter-rater reliability. The Cohen's kappa coefficient was 0.71. On validating the sentiments generated with the TextBlob library, we found that they matched with 84\% and 81\% of the two sets of manually labeled sentiments.

The tweets can contain the title of the research articles. To check the impact of the article's title in the tweets for tweet sentiment prediction, we performed the above-mentioned machine learning techniques with and without the tweets that had the title of the research articles. We performed word sequence matching on the tweets to check if they have the title of the research article and removed the tweets that had 70\% or more sequence matches. We found that 105,834 articles had tweets that did not match with the title of the article. We further refer to the dataset with 148,712 articles (including tweets with article's title) as \textit{dataset A} and the dataset with 105,834 articles (excluding tweets with article's title) as \textit{dataset B}.
\section{Results}
\subsection{Classification models}

To create a model capable of predicting the binary sentiment as positive or negative, we decided to use NLTK VADER and TextBlob. We used these two libraries to obtain the sentiment score between -1 and 1 for the article title, abstract, and tweet (target variable). As the scores were in the range of -1 and 1, all the scores were transformed as positive, neutral, or negative sentiments (Table \ref{tab:seg-scores}).

\begin{table*}
\centering
  \caption{Segregation of sentiments score.}
  \label{tab:seg-scores}
  \begin{adjustbox}{center}
  \begin{tabular}{cc}
    \toprule
    
        Score range & Sentiment \\
            \midrule
        \![-1,0) & Negative \\
        0 & Neutral \\
         (0,1\!] & Positive \\
     \bottomrule
  \end{tabular}
  \end{adjustbox}
\end{table*}

A research article may have multiple tweets related to it, and these tweets may differ from each other in terms of the sentiments expressed. For any given article, we calculated the mean and median of all the sentiments expressed in tweets about it. Table \ref{tab:sent-label-assign} shows some examples with tweets from the dataset to demonstrate the assignment of a sentiment label using the mean of tweets’ sentiment.

\begin{table*}
\centering
\caption{Examples of sentiment label assignment.}
\label{tab:sent-label-assign}
\begin{adjustbox}{center}
\begin{tabular}{ llllll }
 \toprule
    
 \textbf{Article} & 
 \textbf{\vtop{\hbox{\strut 1st Tweet and}\hbox{\strut Sentiment}}} &
 \textbf{\vtop{\hbox{\strut 2nd Tweet and}\hbox{\strut Sentiment}}} &
 \textbf{\vtop{\hbox{\strut 3rd Tweet and}\hbox{\strut Sentiment}}} &
 \textbf{\vtop{
 \hbox{\strut Mean of }
 \hbox{\strut tweets' }
 \hbox{\strut sentiment}}
 } &

 \textbf{\vtop
 {
 \hbox{\strut Final}
 \hbox{\strut sentiment}
 \hbox{\strut class label}
 }
 } 
 
 \\
 \midrule
 
 Article 1 & \vtop{\hbox{\strut Researchers in } \hbox{\strut Norway }
 \hbox{\strut investigate}
 \hbox{\strut mortality risk}\hbox{\strut of individuals}\hbox{\strut  after the death }\hbox{\strut of a spouse}
 \hbox{\strut (-0.7184)}}
 & 
 \vtop{
 \hbox{\strut Can you die of a} \hbox{\strut  broken heart?}
 \hbox{\strut If your spouse}
 \hbox{\strut dies, your death}
 \hbox{\strut risk substantially}
 \hbox{\strut increases }
 \hbox{\strut (-0.9186)}
 }
 
 &
\vtop{
 \hbox{\strut A sad study:} 
\hbox{\strut spouses much }
\hbox{\strut more likely to }
\hbox{\strut die after being}
\hbox{\strut widowed}
\hbox{\strut (-0.885)} 
 }

& -0.8407 & Negative 
\\
\\
Article 2 &
\vtop{
 \hbox{\strut Presentation} 
 \hbox{\strut of the ABC}
 \hbox{\strut Best Paper}
 \hbox{\strut Award 2013 to}
 \hbox{\strut Sherrie Elzey.}
 \hbox{\strut Read the }
 \hbox{\strut winning paper}
 \hbox{\strut (0.9022)}
 }
&

\vtop{
 \hbox{\strut ABC Best} 
 \hbox{\strut Paper Award}
 \hbox{\strut 2013 goes to}
 \hbox{\strut lead authors}
 \hbox{\strut Sherrie Elzey}
 \hbox{\strut and De-Hao }
 \hbox{\strut Tsai. Read }
 \hbox{\strut their article }
 \hbox{\strut for free}
 \hbox{\strut (0.9001)}
 }
& NA & 0.90115 & Positive 
\\
\\
Article 3 &
\vtop{
 \hbox{\strut Latest article} 
 \hbox{\strut from our }
 \hbox{\strut research team}
 \hbox{\strut has been }
 \hbox{\strut published} 
 \hbox{\strut about using}
 \hbox{\strut  School }
 \hbox{\strut  Function}
 \hbox{\strut Assessment!}
 \hbox{\strut (0)}
 }

&

\vtop{
 \hbox{\strut Article on} 
 \hbox{\strut  using}
 \hbox{\strut  School }
 \hbox{\strut  Function}
 \hbox{\strut Assessment} 
 \hbox{\strut now online}
 \hbox{\strut (0)}
 }

& NA & 0 & Neutral \\

\bottomrule
\end{tabular}
\end{adjustbox}
\end{table*}

A summary of the variations of the articles tweets’ sentiments is shown in Tables \ref{tab:cases-sent-twitter} and \ref{tab:cases-sent-twitter-no-title}. In all four cases mentioned, the sentiment library and the metric for multiple sentiments changed while the data was the same. Table \ref{tab:cases-sent-twitter} shows the variations of sentiments on dataset A and Table \ref{tab:cases-sent-twitter-no-title} shows the variations of sentiments on dataset B.

\begin{table*}[h!]
  \caption{Sentiments on dataset A using different libraries and metrics.}
  \label{tab:cases-sent-twitter}
  \begin{adjustbox}{center}
  \begin{tabular}{llllll}
    \toprule
   
Experiment & 
\vtop{
\hbox{\strut Sentiment}
\hbox{\strut library}
} &
\vtop{
\hbox{\strut Metric for}
\hbox{\strut multiple }
\hbox{\strut sentiments}
} & 
\vtop{
\hbox{\strut Number of}
\hbox{\strut positive}
\hbox{\strut sentiments}
} & 
\vtop{
\hbox{\strut Number of}
\hbox{\strut negative}
\hbox{\strut sentiments}
} & 
\vtop{
\hbox{\strut Number of}
\hbox{\strut neutral}
\hbox{\strut sentiments}
}
\\
    \midrule
    
case 1 & VADER & mean & 55,833 ($\approx 37.5\%$) & 37,957 ($\approx 25.5\%$) & 54,922 ($\approx 36.9\%$)
\\

case 2 & VADER & median & 45,606 ($\approx 30.6\%$) & 32,754 ($\approx 22\%$) & 70,352 ($\approx 47.3\%$)\\

case 3 & TextBlob & mean & 67,035 ($\approx 45\%$) & 16,881 ($\approx 11.3\%$) & 64,796 ($\approx 43.6\%$)\\

case 4 & TextBlob & median & 53,466 ($\approx 36\%$) & 13,748 ($\approx 9.2\%$) & 81,498 ($\approx 54.8\%$)\\

    \bottomrule
  \end{tabular}
  \end{adjustbox}
\end{table*}

\begin{table*}[h!]
  \caption{Sentiments on dataset B using different libraries and metrics.}
  \label{tab:cases-sent-twitter-no-title}
  \begin{adjustbox}{center}
  
  \begin{tabular}{llllll}
    \toprule
   
Experiment & 
\vtop{
\hbox{\strut Sentiment}
\hbox{\strut library}
} &
\vtop{
\hbox{\strut Metric for}
\hbox{\strut multiple }
\hbox{\strut sentiments}
} & 
\vtop{
\hbox{\strut Number of}
\hbox{\strut positive}
\hbox{\strut sentiments}
} & 
\vtop{
\hbox{\strut Number of}
\hbox{\strut negative}
\hbox{\strut sentiments}
} & 
\vtop{
\hbox{\strut Number of}
\hbox{\strut neutral}
\hbox{\strut sentiments}
}
\\
    \midrule
    
case 1 & VADER & mean & 44,866 ($\approx 42.4\%$) & 26,664 ($\approx 25.1\%$) & 34,304 ($\approx 32.4\%$)
\\

case 2 & VADER & median & 38,038 ($\approx 35.9\%$) & 23,124 ($\approx 21.8\%$) & 44,672 ($\approx 42.2\%$)\\

case 3 & TextBlob & mean & 54,169 ($\approx 51.1\%$) & 11,841 ($\approx 11.1\%$) & 39,824 ($\approx 37.6\%$)\\

case 4 & TextBlob & median & 45,254 ($\approx 42.7\%$) & 9,551 ($\approx 9\%$) & 51,029 ($\approx 48.2\%$)\\

    \bottomrule
  \end{tabular}
  \end{adjustbox}
\end{table*}

For classification, we have certain parameters such as
\begin{itemize}
    \item True Negative (TN): Observation is negative and is predicted as negative.
    \item False Positive (FP): Observation is negative but is predicted as positive.
    \item False Negative (FN): Observation is positive but is predicted as negative.
    \item True Positive (TP): Observation is positive and is predicted as positive.
\end{itemize}

The evaluation metrics used for classification models are based on these four mentioned parameters:

\textbf{Accuracy} shows the ratio of correct predictions to total observations.
    \begin{equation}
    Accuracy = \frac{TP + TN}{TP + FP + TN + FN}
    \label{eq:intro_accuracy}
    \end{equation}

\textbf{Precision} is the ratio of the True Positive parameter to the sum of True Positive and False Positive parameters.

    \begin{equation}
    Precision = \frac{TP}{TP + FP}
    \label{eq:intro_precision}
    \end{equation}   

\textbf{Recall} is the ratio of the True Positive parameter to the sum of True Positive and False Negative parameters.
        \begin{equation}
        Recall = \frac{TP}{TP + FN}
        \label{eq:intro_recall}
        \end{equation}

\textbf{F-1 score} is the weighted average of Precision and Recall and derives a balance between Precision and Recall. Unlike accuracy, the F-1 score considers both false positives and false negatives.
          \begin{equation}
          \textit{F-1 score} =
          \frac{2TP}{2TP + FP + FN}
        \label{eq:intro_f1}
        \end{equation}   
        
\textbf{Weighted Average Precision/Recall/F-1 Score} calculates the metrics of each label and finds the average weighted by support (the number of true instances for each label).

\subsubsection{Classification models with two class labels}

Here, we exclude neutral sentiments to consider only the positive and negative sentiments expressed in the tweets. The number of positive and negative tweets differs especially in cases 3 and 4. To overcome this class imbalance, we used the Synthetic Minority Oversampling TEchnique (SMOTE) \citep{chawla2002smote}, which creates synthetic minority class examples. Additionally, we used 10-fold cross-validation and the Grid Search mechanism, which performs an exhaustive search over given parameters to build the best possible model \citep{pedregosa2011scikit}. We built machine learning models with an 80–20 train test split to predict the binary sentiment of the tweets for all four cases for both  datasets A and B (Tables \ref{tab:cases-sent-twitter} and \ref{tab:cases-sent-twitter-no-title}). 
In both scenarios, the case 4 model returned the highest accuracy.

\textbf{Case 4 Experiment:} 

\begin{figure}[hbt!]
\centering
\begin{subfigure}{.5\textwidth}

    \centering
    \includegraphics[width=\textwidth]{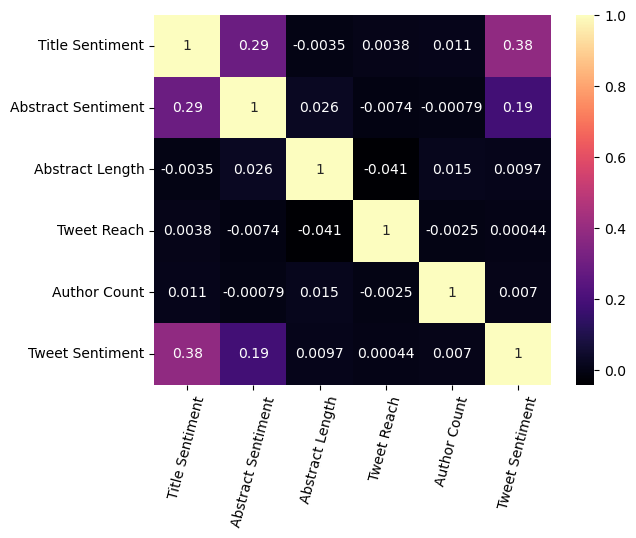}
    \caption{Dataset A: Tweets with article's title.}
    \label{fig:tw-corr-2class-w-title}
\end{subfigure}%
\begin{subfigure}{.5\textwidth}
    \centering
    \includegraphics[width=\textwidth]{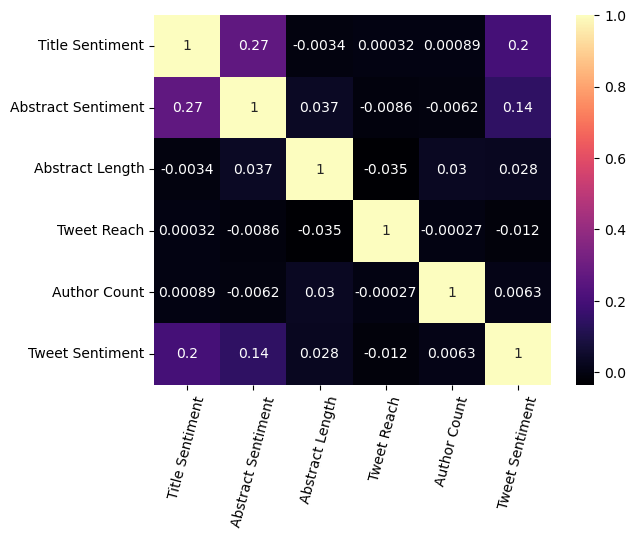}
    \caption{Dataset B: Tweets without article's title.}
    \label{fig:tw-corr-2class-wo-title}
\end{subfigure}
\caption{Correlation matrix of features with two class labels – case 4.}
\label{fig:tw-corr-2class}
\end{figure}

\begin{figure}[hbt!]
\centering
\begin{subfigure}[b]{0.9\textwidth}
   \includegraphics[width=1\linewidth]{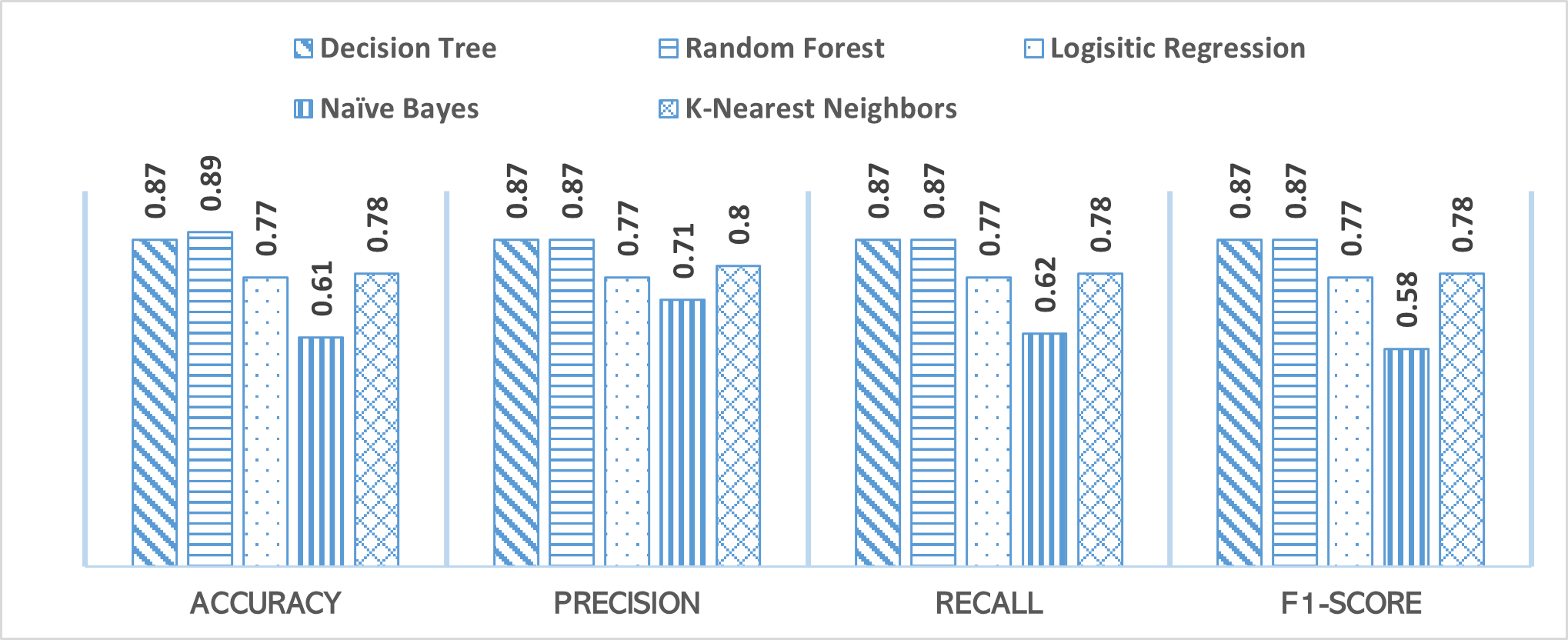}
   \caption{Performance of models on Dataset A: Tweets with article's title.}
   \label{fig:tw-performance-2class-with-title} 
\end{subfigure}

\hspace{5mm}

\begin{subfigure}[b]{0.9\textwidth}
   \includegraphics[width=1\linewidth]{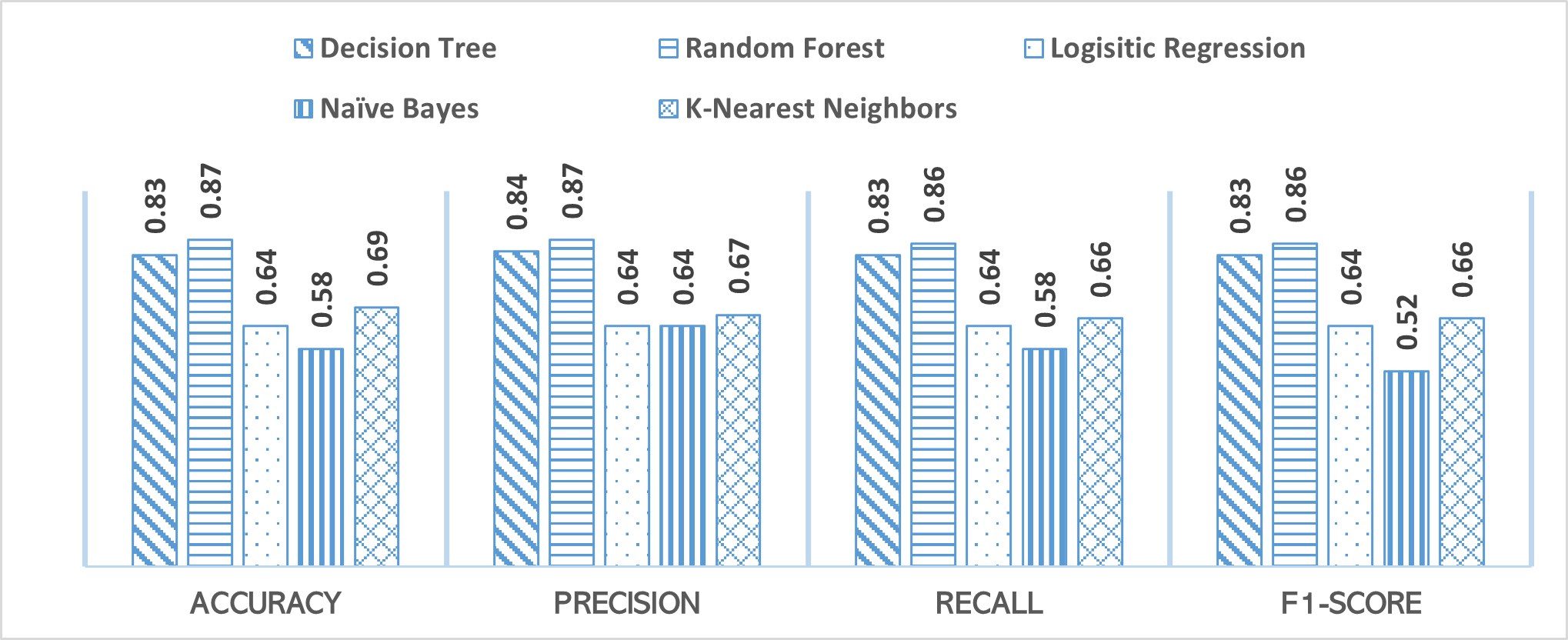}
   \caption{Performance of models on Dataset B: Tweets without article's title.}
   \label{fig:tw-performance-2class-without-title}
\end{subfigure}
\caption{Performance of classification models with two class labels – case 4.}
\label{fig:tw-performance-2class}
\end{figure}

\begin{figure}[hbt!]
\centering
\begin{subfigure}[b]{0.8\textwidth}
   \includegraphics[width=1\linewidth]{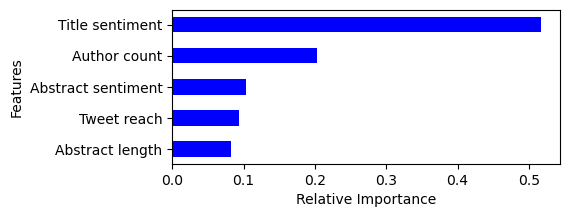}
   \caption{Important features on Dataset A: Tweets with article's title.}
   \label{fig:tw-feat-imp-2class-with-title} 
\end{subfigure}

\hspace{5mm}

\begin{subfigure}[b]{0.8\textwidth}
   \includegraphics[width=1\linewidth]{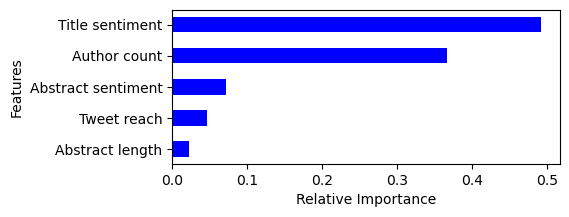}
   \caption{Important features on Dataset B: Tweets without article's title.}
   \label{fig:tw-feat-imp-2class-without-title}
\end{subfigure}
\caption{Important features for two-class label classification.}
\label{fig:tw-feat-imp-2class}
\end{figure}

Here, we used TextBlob as the sentiment library, the median of the sentiments as the metric for multiple tweets to predict whether the tweet sentiments were positive or negative. Figure \ref{fig:tw-corr-2class} shows the correlation matrix for the features used in this experiment for datasets A and B. On comparing the correlation of title sentiment with the sentiment of the tweets in Figures \ref{fig:tw-corr-2class-w-title} and \ref{fig:tw-corr-2class-wo-title}, we observed a significant drop in the correlation on dataset B. Figure \ref{fig:tw-performance-2class} shows a comparison of the performance of machine learning models in terms of accuracy and weighted average scores for precision, recall and F-1. The Random Forest and Decision Tree models performed better than the other models for both datasets A and B. From Figures \ref{fig:tw-performance-2class-with-title} and \ref{fig:tw-performance-2class-without-title}, we observed that the performance of Decision Tree and Random Forest models did not vary much, but there was a significant performance variation for Logistic Regression and K-Nearest Neighbors models. The important features for the binary classification derived from the Random Forest model are shown in Figure \ref{fig:tw-feat-imp-2class}. The sentiment of the title of the research article is the most important feature, followed by the author count on both datasets A and B. From Figures \ref{fig:tw-feat-imp-2class-with-title} and \ref{fig:tw-feat-imp-2class-without-title}, we observe that irrespective of the title being in the tweet, the title's sentiment is still the most important feature for predicting the tweet sentiment. Table \ref{tab:other-perf-2labels} shows the best results for cases 1–3 in predicting positive and negative tweet sentiments.

\begin{table*}[hbt!]
\centering
  \caption{Best results for cases 1–3 with two-class labels.}
  \label{tab:other-perf-2labels}
  \begin{adjustbox}{center}
  
  \begin{tabular}{cccc}
    \toprule
    \multicolumn{4}{c}{Dataset A: Tweets with article's titles}\\
    \midrule
        Case Number & Model & Accuracy & F-1 Score \\
            \midrule
       1 & Random Forest & 0.81 & 0.81 \\ 
       2 & Random Forest & 0.83 & 0.83 \\ 
       3 & Random Forest & 0.85 & 0.85 \\

    \midrule
    \multicolumn{4}{c}{Dataset B: Tweets without article's titles}\\
    \midrule
        Case Number & Model & Accuracy & F-1 Score \\
            \midrule
       1 & Random Forest & 0.77 & 0.76 \\ 
       2 & Random Forest & 0.78 & 0.78 \\ 
       3 & Random Forest & 0.85 & 0.80 \\ 
     \bottomrule
  \end{tabular}
  \end{adjustbox}
\end{table*}

\subsubsection{Classification models with three class labels}

From Tables \ref{tab:cases-sent-twitter} and \ref{tab:cases-sent-twitter-no-title}, we can observe that the neutral sentiments are numerous—a point that cannot be neglected in the analysis. For this experiment, the sentiment of a research article has three possible classes. Here, we also used SMOTE to deal with class-imbalance problems by upsampling the minority class label data. Additionally, we used a 10-fold cross-validation and Grid Search mechanism to build the best possible model. We built machine learning models with an 80–20\% train test split for all four cases on datasets A and B (Tables \ref{tab:cases-sent-twitter} and \ref{tab:cases-sent-twitter-no-title}) that predict positive, neutral, and negative tweet sentiments. We observed that the case 4 experiment set-up generated better results than did the set-ups used for other cases on both datasets A and B.

\begin{figure}[hbt!]
\centering
\begin{subfigure}{.5\textwidth}
    \centering
    \includegraphics[width=\textwidth]{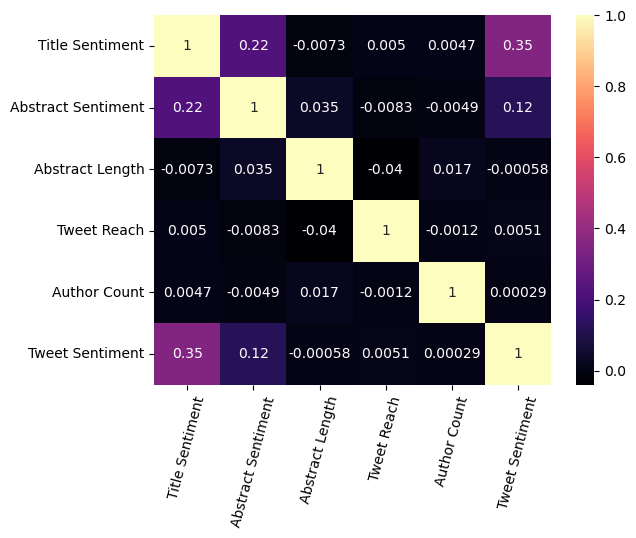}
    \caption{Dataset A: Tweets with article's title.}
    \label{fig:tw-corr-3class-w-title}
\end{subfigure}%
\begin{subfigure}{.5\textwidth}
    \centering
    \includegraphics[width=\textwidth]{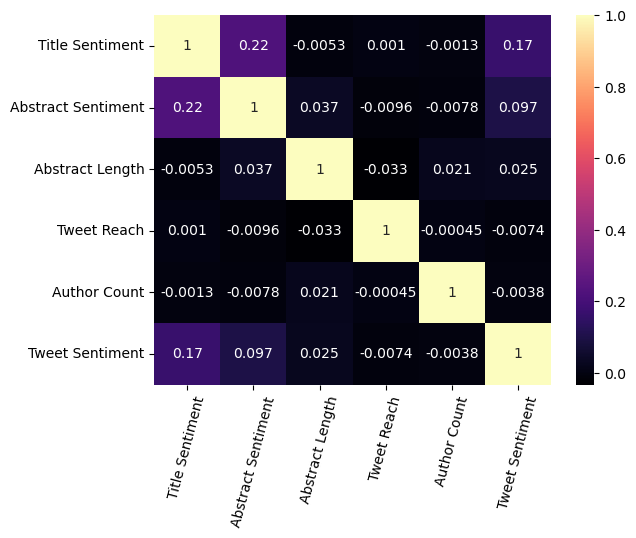}
    \caption{Dataset B: Tweets without article's title.}
    \label{fig:tw-corr-3class-wo-title}
\end{subfigure}
\caption{Correlation matrix of features with three class labels – case 4.}
\label{fig:tw-corr-3class}
\end{figure}

\begin{figure}[hbt!]
\centering
\begin{subfigure}[b]{0.9\textwidth}
   \includegraphics[width=1\linewidth]{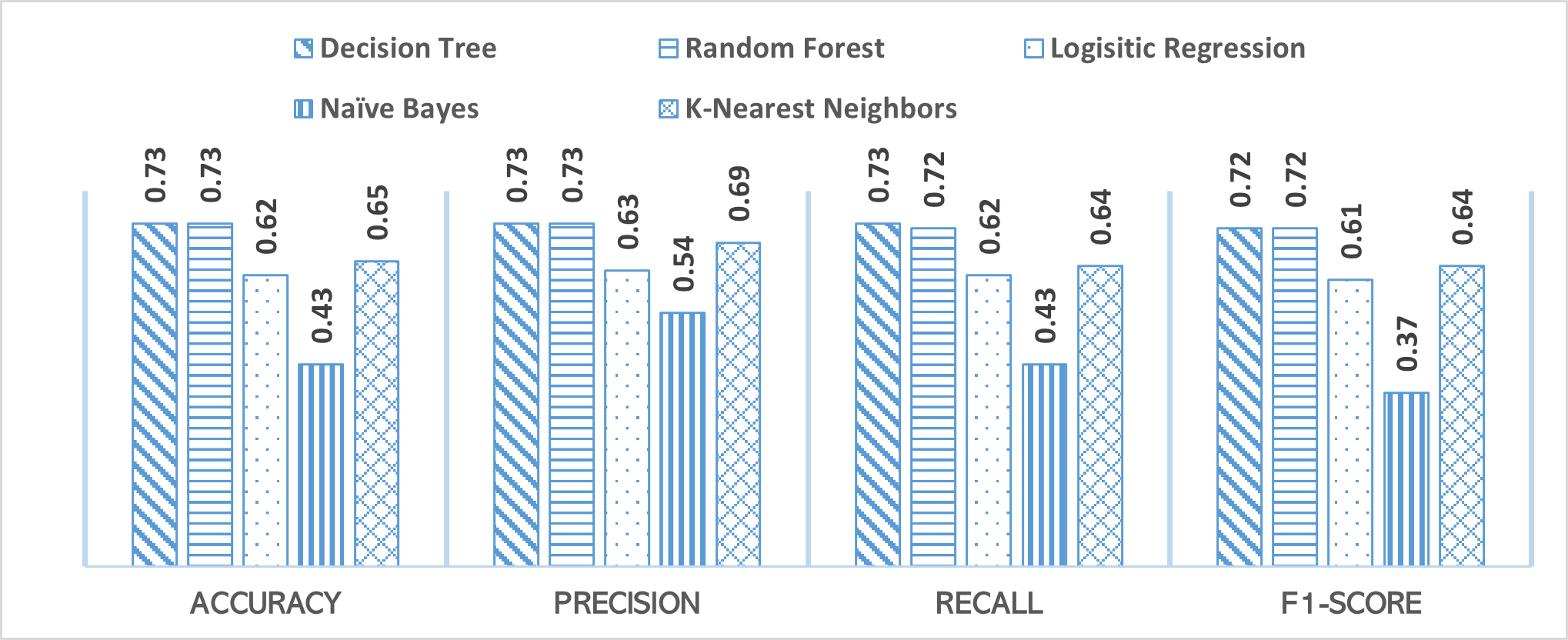}
   \caption{Performance of models on Dataset A: Tweets with article's title.}
   \label{fig:tw-performance-3class-with-title} 
\end{subfigure}

\hspace{5mm}

\begin{subfigure}[b]{0.9\textwidth}
   \includegraphics[width=1\linewidth]{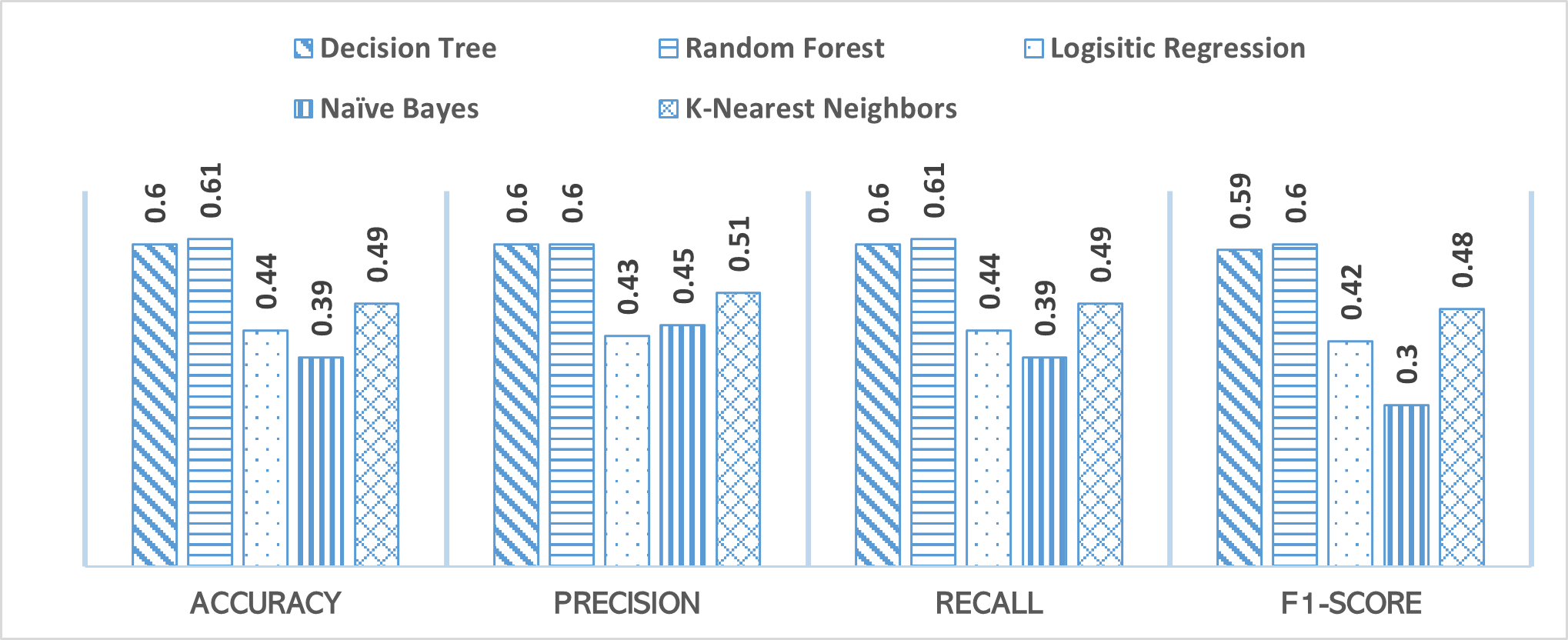}
   \caption{Performance of models on Dataset B: Tweets without article's title.}
   \label{fig:tw-performance-3class-without-title}
\end{subfigure}
\caption{Performance of classification models with three class labels – case 4.}
\label{fig:tw-performance-3class}
\end{figure}

\begin{figure}[hbt!]
\centering
\begin{subfigure}[b]{0.8\textwidth}
   \includegraphics[width=1\linewidth]{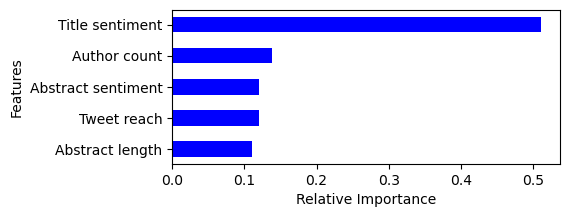}
   \caption{Important features on Dataset A: Tweets with article's title.}
   \label{fig:tw-feat-imp-3class-with-title} 
\end{subfigure}

\hspace{5mm}

\begin{subfigure}[b]{0.8\textwidth}
   \includegraphics[width=1\linewidth]{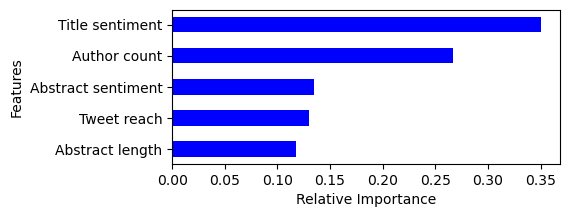}
   \caption{Important features on Dataset B: Tweets without article's title.}
   \label{fig:tw-feat-imp-3class-without-title}
\end{subfigure}
\caption{Important features for three-class label classification.}
\label{fig:tw-feat-imp-3class}
\end{figure}

\textbf{Case 4 Experiment:} Here, we used TextBlob as the sentiment library, the median of the sentiments as the metric for multiple tweets to predict whether the tweet sentiments were positive, neutral, or negative. Figure \ref{fig:tw-corr-3class} shows the correlation matrix for the features used in this experiment for datasets A and B. We observed that the title sentiment of the research article had the highest correlation with the tweet sentiment. From Figures \ref{fig:tw-corr-3class-w-title} and \ref{fig:tw-corr-3class-wo-title}, we observed that the correlation between title sentiment and tweet sentiment decreased by almost half from dataset A to B. Figure \ref{fig:tw-performance-3class} shows the performance of the different models in terms of accuracy and weighted average scores for precision, recall, and F-1. Again, the tree-based models performed better than the other models on datasets A and B. Naive Bayes was the worst model. From Figures \ref{fig:tw-performance-3class-with-title} and \ref{fig:tw-performance-3class-without-title}, we observed that the performance of the models significantly dropped compared with the previous two class classification. Surprisingly, the Naive Bayes model performed better on dataset B. Figure \ref{fig:tw-feat-imp-3class} shows the important features derived from the Random Forest model. We observed that the sentiment of the research article’s title is a vital feature in both datasets A and B. However, on comparing Figures \ref{fig:tw-feat-imp-3class-with-title} and \ref{fig:tw-feat-imp-3class-without-title}, we observed that the importance of title sentiment decreases by about 15\% when the model is built using the tweets that do not contain the article's title. Table \ref{tab:other-perf-3labels} shows the best results for cases 1–3 in predicting positive, neutral, and negative tweet sentiments. Here, we observe that cases 1 and 2 that build the models on the VADER sentiment library did not exceed 50\% for accuracy and F-1 score.

\begin{table*}[hbt!]
\centering
  \caption{Best results for cases 1–3 with three labels.}
  \label{tab:other-perf-3labels}
  \begin{adjustbox}{center}
  \begin{tabular}{cccc}
    \toprule
    \multicolumn{4}{c}{Dataset A: Tweets with article's titles}\\
    \midrule
        Case Number & Model & Accuracy & F-1 Score \\
            \midrule
       1 & Random Forest & 0.46 & 0.46 \\ 
       2 & Random Forest & 0.49 & 0.45 \\ 
       3 & Random Forest & 0.68 & 0.66 \\

    \midrule
    \multicolumn{4}{c}{Dataset B: Tweets without article's titles}\\
    \midrule
        Case Number & Model & Accuracy & F-1 Score \\
            \midrule
       1 & Random Forest & 0.46 & 0.45 \\ 
       2 & Random Forest & 0.47 & 0.44 \\ 
       3 & Random Forest & 0.56 & 0.56 \\ 
     \bottomrule
  \end{tabular}
  \end{adjustbox}
\end{table*}

\subsection{Regression models}
We implemented some regression models to predict the sentiment score of the tweet. We used 80\% training and 20\% testing split and obtained the results shown in Table \ref{tab:reg-results}. The regression models did not perform well based on the R-squared values.

\begin{table*}[hbt!]
  \caption{Results of the regression models.}
  \label{tab:reg-results}
  \begin{adjustbox}{center}
  \begin{tabular}{ccc}
    \toprule

\multicolumn{3}{c}{Dataset A: Tweets with article's titles}\\
\midrule

Model & Mean Squared Error & R-Squared \\

\midrule

Multiple Linear Regression & 0.091 & 0.008
\\
Decision Tree & 0.189 & -1.051
\\
Random Forest & 0.104 & -0.130
\\
Support Vector Regression & 0.093 & -0.014
\\

\midrule
\multicolumn{3}{c}{Dataset B: Tweets without article's titles}\\
\midrule

Model & Mean Squared Error & R-Squared \\

\midrule

Multiple Linear Regression & 0.104 & 0.006
\\
Decision Tree & 0.470 & -1.095
\\
Random Forest & 0.119 & -0.133
\\
Support Vector Regression & 0.106 & -0.009
\\

     \bottomrule
  \end{tabular}
  \end{adjustbox}
\end{table*}
\section{Discussion}
In this study, to predict the tweet sentiments for research articles, we used the NLTK VADER and TextBlob libraries to obtain sentiment scores. We built machine learning models to predict the sentiment as a binary class label (positive or negative) and as three class labels (positive, neutral, or negative). We also used the Stanford CoreNLP library for a sample of 7,000 articles. However, this library resulted in about 95\% of the sentiments being neutral. We observed that the dataset with sentiments generated using the TextBlob library had lower correlations among the title sentiment, abstract sentiment, and tweet sentiment when compared to those correlations on the dataset having sentiments generated using NLTK VADER library.

Several studies that analyzed sentiments of tweets for scholarly articles have shown that the majority of the tweets for altmetrics have no sentiment value (neutral), and positive tweets are more than negative tweets. \cite{thelwall2013tweeting} analyzed 270 tweets for papers published in 2012 and manually labeled 96\% neutral, 4\% positive, 0\% negative. \cite{friedrich2015adapting} performed sentiment analysis on 1,000 tweets using the SentiStrength tool and found that 94.8\% of tweets were neutral, 4.3\% positive, and 0.9\% negative.  \cite{friedrich2015tweets} analyzed 487,610 tweets using SentiStrength and found 81.7\% neutral, 11\% positive, and 7.3\% negative. Our study also shows that sentiments of tweets are more neutral than positive or negative. By using NLTK VADER and TextBlob libraries, we avoided the need to manually label tweets and have a lower level of neutral sentiments (43.6\%), which reduces the class imbalance issue to an extent. Additionally, we used SMOTE to overcome class imbalance issues.

Any given research article could be discussed in multiple tweets. Therefore, we evaluated the final sentiment of all the tweets related to each given research article using the mean and median of the sentiments. However, we found that the mean and median did not matter much for the prediction of the tweet sentiment. 

\cite{raamkumar2018investigating} studied tweets of computer science research articles using the Microsoft academic graph (MAG) dataset and found that the research impact (in terms of bibliometrics and altmetrics) of papers that had the three sentiments were better than those having just neutral sentiment. Interestingly, their study also used the TextBlob library to classify 49,849 tweets related to 12,967 computer science research papers. They found that  97.16\% of the tweets had neutral sentiments, 2.8\% had positive, and 0.05\% had negative sentiments. Compared to our study, we observe that although the dataset is different, the library for sentiment classification is the same and tweets are still about the scholarly articles. The discrepancy in the percentage of the three sentiments could be because of the scientific field selection. Various fields might have various sentiment values for tweets. In computer science, users might just post the article with the given title without giving any reactions. However, most of our papers are health-related, which could explain the interest and increase in non-neutral sentiments.  

In this study, we observed that the tree-based classifiers generally performed better than the other classifiers did. The best model we built was Random Forest using a median of tweets and the TextBlob library (case 4). When the tweets having the research article's title were included when building the model, it generated 89\% accuracy for the binary classification and 73\% accuracy for the three-label classification. When those tweets were excluded when building the model, it generated 73\% accuracy for binary classification and 61\% accuracy for the three-label classification.  The most important feature in all the models was the sentiment of the research article’s title.

We built regression models to determine the exact sentiment value for a tweet in a range of -1 (highly negative) to +1 (highly positive). However, these results were not as good as those from the classification with two or three class labels. Thus, we found it is difficult to determine the exact sentiment as a continuous value. However, in general, we were able to determine whether a tweet expressed a positive, negative, or neutral sentiment toward a given research article.

Our work has some limitations, such as considering several research disciplines that might affect the sentiments. For example, some disciplines might share similar keywords in the title and abstract but receive different sentiments. Additionally, we did not consider the impact of the publisher highlights on the tweet sentiment. Further, tweet sentiment may reflect a variety of factors but are not necessarily related directly to the specifics of the paper. The tweet sentiment, by no means, is an accurate indicator of the reaction to the research paper content. For example, we have no easy way of knowing if the tweeter read the study or just replied to the title or abstract. Furthermore, some users might have positive or negative sentiments toward scientific topics in general (e.g., pro- and anti-vaccination views) and not necessarily the scientific paper they are tweeting about. However, the sentiment of tweets is a significant metric for comprehending what the tweeters feel about the research. In a broader sense, this contributes to our understanding of the societal impact of research.

\section{Conclusion and Future Work}

In this study, we analyzed tweets related to research articles. We observed various patterns for the number of tweets, the research articles’ subjects, and the publication years of the research articles. Due to the fact that tweets often include the title of the research paper, we created two distinct datasets: one with the title of the article in the tweets and one without. We built machine learning models to predict the sentiment of the tweet for a given research article. We developed classification models for two class and three class labels (positive, neutral, or negative). The tree-based classifiers generally outperformed other classifiers. We discovered that the sentiment expressed in the research paper title was the most significant feature in all models, followed by the number of authors. In future research, we plan to use neural networks to develop more robust machine learning models and provide tailored recommendations \citep{alhoori2016, alhoori2017} to authors. We also plan to expand the research area using word2vec models and extract important features from the title and the abstract of the research article that contribute to predicting the sentiments expressed in tweets.

\section*{Acknowledgments}
This work is supported in part by NSF Grant No. 2022443.

\bibliography{Main}

\end{document}